# Contextual Risk and Its Relevance in Economics

Diederik Aerts and Sandro Sozzo
Center Leo Apostel for Interdisciplinary Studies
Brussels Free University, Pleinlaan 2, 1050 Brussels
E-Mails: diraerts@vub.ac.be, ssozzo@vub.ac.be

**Abstract**

Uncertainty in economics still poses some fundamental problems illustrated, e.g., by the *Allais* and *Ellsberg paradoxes*. To overcome these difficulties, economists have introduced an interesting distinction between *risk* and *ambiguity* depending on the existence of a (classical Kolmogorovian) probabilistic structure modeling these uncertainty situations. On the other hand, evidence of everyday life suggests that context plays a fundamental role in human decisions under uncertainty. Moreover, it is well known from physics that any probabilistic structure modeling contextual interactions between entities structurally needs a non-Kolmogorovian quantum-like framework. In this paper we introduce the notion of *contextual risk* with the aim of modeling a substantial part of the situations in which usually only ambiguity is present. More precisely, we firstly introduce the essentials of an operational formalism called the *hidden measurement approach* in which probability is introduced as a consequence of fluctuations in the interaction between entities and contexts. Within the hidden measurement approach we propose a *sphere model* as a mathematical tool for situations in which contextual risk occurs. We show that a probabilistic model of this kind is necessarily non-Kolmogorovian, hence it requires either the formalism of quantum mechanics or a generalization of it. This insight is relevant, for it explains the presence of quantum or, better, quantum-like, structures in economics, as suggested by some authors, and can serve to solve the aforementioned paradoxes.

## 1. The notion of contextual risk

The *expected utility hypothesis* requires that individuals evaluate uncertain prospects according to their expected level of 'satisfaction' or 'utility' (von Neumann and Morgenstern, 1944). This hypothesis characterizes the predominant model of choice under uncertainty in economics. However, it is well known that examples and experiments exist in the literature, e.g., the *Allais* and *Ellsberg paradoxes* (Allais 1953, Ellsberg 1961), which show that real-life subjects seem to prefer 'sure choices' over 'choices containing ambiguities', which entails a violation of the expected utility hypothesis. Various attempts have been elaborated since the fifties to understand and explain these paradoxes, but a universally accepted interpretation is still an unachieved goal.

Frank Knight introduced a distinction between different kinds of uncertainty (Knight 1921), and Daniel Ellsberg inquired into the conceptual differences between them (Ellsberg 1961). More explicitly, Ellsberg put forward the notion of *ambiguity* as an uncertainty without any well-defined probability measure to model this uncertainty, as opposed to *risk*, where such a probability measure does exist. We can understand the difference between ambiguity and risk by considering the situation introduced by Ellsberg himself. Two urns are considered, one of them containing 30 red balls, and the other containing 60 balls that are that are either black or yellow, the latter in unknown proportion. Two bets are introduced, bet I 'betting on red', and bet II 'betting on black'. Bet I concerns risk, since the probability involved is known, namely 1/3 to win and 2/3 to loose. For bet II however, since it is only known that the sum of the black and the yellow balls is 60, the number of black balls is not explicitly known. This means that if no additional information is given specifying in more detail the situation, bet II is related to a situation of ambiguity. We do not consider further the Ellsberg paradox here, but refer to (Aerts and Sozzo 2011) where we analyze it in detail.

It is important to note that the probability description meant by Ellsberg and economists, when they define 'risk' as 'uncertainty for which an explicit probability description is known', is a classical probability description, in the sense that is satisfies Kolmogorov's axioms (Kolmogorov 1933), hence it is commonly called a *Kolmogorovian probability*. It is well known that the probability description used in quantum physics is not Kolmogorovian (Accardi 1982, Pitowsky 1989, Beltrametti 2001). It also has been understood where quantum probability differs from classical Kolmogorovian probability. Namely, a classical Kolmogorovian probability can only model a situation of 'lack of knowledge about an underlying deterministic reality' and such that the tests involved to measure the probabilities do not influence the underlying situation in a non-determined way. This means that if 'contextual influence' is present, in the sense that 'tests do influence the underlying situation a non-determined way', such a situation cannot be described by a classical Kolmogorovian probability. The probability appearing in quantum mechanics can exactly describe the presence of such contextual influence.

Now, contextual influence commonly appears in situations of risk. Let us consider, for example, the risk of having an accident. It is evident that if a person is in a context where he or she is sitting in a chair reading the newspaper on his or her terrace, the risk of having an accident is low if compared with the risk of getting an accident when this person is in the context of sitting in a car next to a reckless driver. In case 'risk to have an accident' is considered with respect to this person, then the two mentioned contexts will have a different influence on the probabilities describing this risk. Similar examples can be found in risk management, where one has to identify, monitor and control the external factors, including accidents, natural causes and disasters, which can potentially affect given financial operations.

The above considerations bring us to the first aim of the present article. Namely, we want to introduce the notion of *contextual risk*, as a generalization of the notion of risk as conceived in traditional economics. *Contextual risk* is uncertainty that is present in a situation for which an explicit probabilistic description exists, 'but' the probability used can be non classical, i.e. non-Kolmogovian. Possibly the probability can be of a quantum nature, but this is not necessarily so. We will indeed see in next section that more general contextual probability theories exist than the quantum one, and also these will be considered for the notion of contextual risk. The second aim of this article is to point out that with this notion of contextual risk, we are able to model uncertainty that traditionally is classified as ambiguity. More specifically, we are able to model the ambiguity appearing in the Ellsberg paradox (Aerts and Sozzo 2011). In Sec. 3 we present a simple example that can be used to model situations in which contextual risk occurs.

## 2. The hidden measurement approach as a contextual probability theory

One of the characteristic traits of quantum mechanics is the *measurement context* provoking an indeterministic influence on the physical system that is considered. The mathematical formalism of quantum mechanics describes precisely this influence and its corresponding probabilities, which is why the quantum probability model is a probability model which can incorporate the effect of context. The situation of contextual influence outside physics is, in general, more complex than the one encountered in the microscopic world, but generalizations of the mathematical formalism of quantum mechanics can be used in these cases. In particular, in our Brussels research group we have developed a generalization of the contextual probability model employed in quantum physics, which we have called 'the hidden measurement approach' (Aerts 1993, 1998). The development of this hidden measurement approach followed from our finding that the non-Kolmogorovian nature of the quantum probability model is due to a lack of knowledge concerning how context interacts with the system under consideration, i.e. by the presence of *fluctuations* in the interaction between context and system (Aerts 1993, 1995, 1998, 2002, Aerts et al. 1997). Even if we were to suppose that at the ontological level the interaction between context and system engenders a change of state that is deterministic, a lack of knowledge about this interaction gives rise to a probability model that does not satisfy Kolmogorov's axioms. The quantum probability model is of this nature, i.e. can be explained as a model with lack of knowledge of 'how the measurement acts and influences the physical system under study'. Let us illustrate more in detail this hidden measurement approach and how we can apply it to deliver a mathematical theory for the notion of contextual risk that we introduced in this article.

Consider a physical system $S$ and states $p,q,r,...$ that represent different situations in which $S$ can be. Consider measurements $e, f, g,...$ that can be performed on this physical system $S$ being in one of the considered states. The case we consider is such that the outcome of an arbitrary measurement $e$ on the system $S$ in a state $p$ is in general 'not' determined, i.e. we are in a situation where 'uncertainty' is present. More concretely, at least for the case of physics, this means that if we repeat the measurement $e$ on the system $S$ being each time prepared in the same state $p$, generally different outcomes will occur. Let us recall that this does not necessarily mean that 'all considered situations are non-deterministic'. There might well be some states and some measurements giving rise to deterministic outcomes. This is, by the way, also the case in quantum mechanics, and such states with deterministic outcomes are called 'eigenstates' of the measurement in question. Also more determinism than what is allowed in quantum mechanics is possible within the hidden measurement approach. Indeed, when we stated in Sec. 1 that the hidden measurement approach is a generalization of quantum mechanics, this is true, but at the same time the hidden measurement approach is also a generalization of classical mechanics, which is the reason that also the full deterministic case can occur. Both quantum mechanics and classical mechanics are special cases of our approach.

The principal idea of the hidden measurement approach is that the uncertainty can have its origin in two distinct ways. First of all, it is possible that we lack knowledge about the state $p$ of the system. This type of uncertainty can be treated by classical mechanical theories, technically it means that the state $p$ in question is a mixed state and not a pure state, which it would be in case we do not lack knowledge about it. The second possibility is the new one, namely that we lack knowledge about the measurement context $e$, more specifically about 'how the measurement $e$ interacts with the system', or, with other words, 'how the system behaves specifically in the considered measurement context $e$ during the act of measurement'. The uncertainty finding its origin in this 'contextual lack of knowledge about the measurement context' cannot be described by classical mechanics theories. A quantum-like theory is needed, and the hidden measurement approach developed in our Brussels research group is a mathematical theory able to describe all uncertainties of the first 'and' of the second type, hence whether they find their origin in a lack of knowledge about the state $p$ of the system $S$ or in a lack of knowledge about how the measurement $e$ interacts with the system $S$ during the happening of this measurement.

Let us now explain how the hidden measurement approach can be applied to provide a mathematical modeling for the notion of contextual risk. Economists are interested in 'decisions taken by humans with respect to specific situations', such as the situation described in the Ellsberg paradox. In the process of decision making there will generally be a 'cognitive influence' having its origin in the way the mind(s) of the person(s) involved in the decision making relate to the situation that is the subject of the decision making. The role played in physics by 'the physical system' is played in economics by 'the considered situation', the role played in physics by the measurement is played in economics by the 'cognitive context', and the role played in physics by the interaction during the measurement is played in economics by the decision interaction between mind and situation. The cognitive context incorporates in principle all types of cognitive aspects that are able to influence the decision interaction. Quite obviously we generally are in a situation that there is lack of knowledge about the situation itself, but also lack of knowledge about the decision context and how it interacts with the situation. This presence of the specific double lack of knowledge is what makes contextual probability, i.e. the hidden measurement approach, apt for providing a faithful description.

We have been employing already the description by means of contextual probability models to the situation of human decision making a long time ago for the study of an opinion poll (Aerts and Aerts 1994). Of course, this was in the realm of psychology-sociology and at that time we were not aware that our modeling of the opinion poll contains the roots for also modeling the human decisions in economic situations. The years after this initial use of the quantum formalism for describing contextual aspects of human thought we went to investigating the structure of concepts and their combinations (Aerts and Gabora 2005a,b, Gabora and Aerts 2002). This is what made us introduce the notion of 'conceptual landscape' indicating the aspect of human thought that constitutes the cognitive context when decisions are made. More specifically we put forward the hypothesis that human thought consists of two superposed layers: (i) a layer given form by an underlying classical deterministic process, incorporating essentially logical thought and its indeterministic version modeled by classical probability theory; (ii) a layer given form under influence of the totality of the surrounding conceptual landscape, where the different concepts figure as individual entities rather than (logical) combinations of others, with measurable quantities such as 'typicality', 'membership', 'representativeness', 'similarity', 'applicability', 'preference' or 'utility' carrying the influences. We have called the process in this second layer 'quantum conceptual thought', which is indeterministic in essence, and contains holistic aspects, but is equally well, although very differently, organized than logical thought. A substantial part of the 'quantum conceptual thought process' can be modeled by quantum mechanical probabilistic and mathematical structures (Aerts 2009, 2010, Aerts and D'Hooghe 2009).

Before we put forward a simple example of a contextual probability situation in Sec. 3, we want to mention that the line of research consisting of employing the quantum mechanical probability model to describe aspect of human thought has enjoyed meanwhile an explosive growth internationally, and has now become a new emergent domain of research called *Quantum Cognition*. To date, four international conferences have been organized, a fifth one being due in June (Bruza et al. 2007, 2008, 2009, 2010). There is a Wikipedia page explaining the essentials of Quantum Cognition at http://en.wikipedia.org/wiki/Quantum_cognition, where also an overview of the literature can be found. We stress, to conclude this section, that the presence of a quantum structure in cognition does not necessarily entail the requirement that microscopic quantum processes occur in human mind. In fact, we avoid engaging ourselves in such a compelling assumption in the following.

## 3. The sphere model as an example of a contextual probability situation

We introduce in this section a simple example that illustrates the way in which a contextual probability model can be built. We call this example the *sphere model*, since we have used it before under this naming in our study of the hidden measurement approach (Aerts 1993, 1995, 1998 Aerts et al. 1997). The example shows explicitly how contextual structures can be worked out which exhibit a quantum-like behavior and are non-Kolmogorovian from a probabilistic point of view. For this reason, our sphere model can be used as a modeling instrument in economic situations where contextual risk occurs, and we will use it to describe a contextual risk analysis of the Ellsberg paradox situation (Aerts and Sozzo 2011). Let us start with a brief presentation of the model.

The physical system that we consider is a point particle $P$ that can move on the surface of a sphere with center $O$ and radius 1, and we shall denote this surface by $surf$. This particle $P$ is our physical system $S$ (see Fig. 1). In our model of the point particle we consider the unit vector $v$ where the particle is located on $surf$ at a certain instant of time $t$ as representing the state of this particle at time $t$, its place on the surface of the sphere, that we shall denote by $p_v$. Let us now introduce the measurement contexts. We consider two diametrically opposite points $u$ and $-u$ on the surface of the sphere, and an elastic band attached between these two points (see Fig. 1).

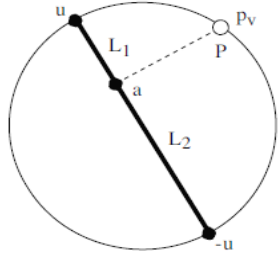

Fig. 1: The sphere model as an example of a contextual probability situation

We shall systematically denote by $[-u,u]$ the 'interval' of real numbers $[-1,1]_u$, coordinating the points of the elastic band between $u$ and $-u$ in such a way that $-1$ coordinates $-u$ and $1$ coordinates $u$. The measurement context $e_u$ consists of the following happening: the particle $P$ falls from its original place $v$ orthogonally onto the elastic band between $u$ and $-u$, and sticks on it in a point $a$ (see Fig. 1), coordinated in the interval $[-u,u]$ by the real number $v \cdot u$. Then the elastic band breaks in a certain point. The breaking of the elastic band is the 'lack of knowledge about the context' that we introduce in the example. Indeed, depending on 'in which point the elastic band breaks' two different changes happen to the state $p_v$ of the system $S$, which is our point particle, now sticking in point $a$ to the elastic band. If the band breaks in a point contained in region $L_1$ from $a$ to $u$, the particle $P$ is pulled towards $-u$, and finally sticks to the sphere again, at point $-u$. This means that in this case state $p_v$ is transformed into state $p_{-u}$. If the band breaks in the region $L_2$ from $a$ to $-u$, the particle $P$ is pulled towards $u$, and finally sticks to the sphere again, at point $u$. This means that in this case state $p_v$ is transformed into state $p_u$. If the elastic band breaks exactly in point $a$, we suppose that there is probability $1/2$ for the particle to be pulled towards $-u$, and hence state $p_v$ changes into state $p_{-u}$, and probability $1/2$ for the particle to be pulled towards $u$, and hence state $p_v$ changes into state $p_u$. The uncertainty introduced by the breaking of the elastic band gives rise to a contextual probability. Let us propose a mathematical model for this. In the interval $[-u,u]$ we consider a random variable $x$, coordinating the point where the elastic band breaks. The random variable $x$ can be interpreted as describing the (hidden) internal construction of the measuring apparatus, in this case the mechanism of breaking of the elastic band. Different mechanisms of breaking of the elastic band can be considered, and these different mechanisms give rise to classical, quantum or intermediate (quantum-like) situations at a probabilistic level. The different mechanisms can be described by considering a distribution $\rho$ of the variable $x \in [-u,u]$ for each of the mechanisms, where

$$\rho : [-u,u] \to [0,+\infty[ \qquad (1)$$

such that

$$\int_\Omega \rho(x)dx \qquad (2)$$

is the probability that the random variable $x \in \Omega \subset [-u,u]$, i.e. that the elastic band breaks in $\Omega$. We also have

$$\int_{[-u,u]} \rho(x)dx = 1 \qquad (3)$$

which expresses the fact that the elastic band always breaks during a measurement. A measurement context $e_u$ where the random variable $x \in [-u,u]$ is distributed as described by $\rho$ will be called a $\rho$-*measurement* and denoted by $e_\rho^u$.

In Fig. 2 we have represented such a measurement $e_\rho^u$ and drawn the probability distribution function $\rho(x)$ such that the definition interval of this function coincides with the line joining the points $u$ and $-u$. The transition probabilities $\mu_\rho(p_u,e_u,p_v)$ and $\mu_\rho(p_{-u},e_u,p_v)$ that the particle $P$ arrives at point $u$ and $-u$ under the influence of the measurement $e_\rho^u$, are respectively given by

$$\mu_\rho(p_u,e_u,p_v) = \int_{-1}^{v \cdot u} \rho(x)dx$$
$$\mu_\rho(p_{-u},e_u,p_v) = \int_{v \cdot u}^{1} \rho(x)dx \qquad (4)$$

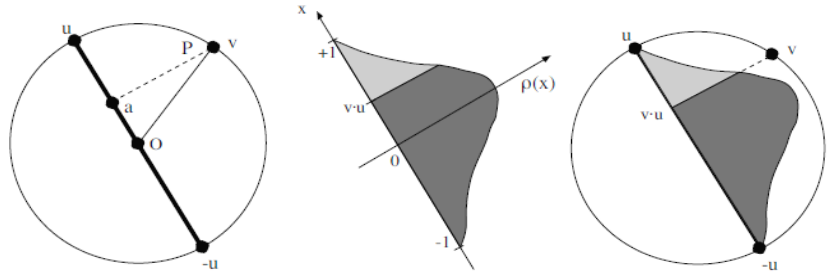

Fig. 2: Different mechanism of breaking of the elastic band

It can be shown that, depending on the particular expression of the density function $\rho$, the sphere model is a model for a spin 1/2 quantum particle (*uniform distribution*) or a model for a deterministic classical system (*deterministic measurements*). But, in general, the sphere model is neither purely classical nor purely quantum (Aerts and Durt 1994). In any case, the contextuality involved in it, together with the fact that the probabilities in Eq. (4) cannot be put into a Kolmogorovian scheme make it a quantum-like model. Indeed, physicists have proven that there is no possibility to capture this type of situation by a unique Kolmogorovian probability model, as it can be seen by using *Pitowsky's classification scheme*, with *polytopes*, or with *Bell-type inequalities* (Accardi 1982, Pitowsky 1989, Beltrametti 2001).

We conclude this section by observing that the sphere model presented above constitutes an example of a contextual structure showing a non-Kolmogorovian quantum-like behavior, hence it can be successfully applied to describe economic situations in which contextual risk appears, thus modeling what has been called `ambiguity' in the literature and supporting the suggestion put forward by some authors according to which quantum-like structures can be identified in economics. We intend to apply the sphere model to a concrete situation in which contextual risk is present, namely the Ellsberg paradox (Aerts and Sozzo 2011).

## 4. Conclusions

Risk has been formally defined in theoretical economics as 'uncertainty under the presence of an explicit mathematical probability model', while ambiguity refers to these situations where no such explicit mathematical probability model exists. These notions were introduced as a consequence of reflections about situations such as the Allais and the Ellsberg paradox. They have become even more important, because it has become clear that the deviations from rational thought for economic agents are linked quite strictly to the presence of ambiguity (see, e.g., Camerer and Weber 1992). Sometimes this is expressed as 'ambiguity aversion', but the situation is more complex. Namely the presence of different types of contexts can provoke deviations from rational thought, and 'aversion of ambiguity' is only one – albeit an important one – of these possible contexts. The traditional definition of risk 'only' takes into account models coming from classical probability theory. From studies in quantum probability it has become clear that it is exactly the presence of context which cannot be modeled within classical probability theory 'and' which can be modeled if recourse is taken to a quantum probability model. This makes it possible for us to introduce the notion of 'contextual risk', and to prove that contextual risk is able to model situations that in the classical economics approach are classified as 'ambiguity'. Hence, our 'contextual risk' approach allows the mathematical modeling of situations of ambiguity by using a quantum probability model. We have already analyzed the original Ellsberg paradox in this 'contextual risk' approach (Aerts and Sozzo 2011), and plan to analyze different known and important situations in economics within our contextual risk approach. A plan on the long term is to investigate 'on which aspects of typical grand economic crises', such as the crisis which started in 2008 and is still continuing, can be shed light – in the form of providing new and insightful modeling – within this contextual risk approach. We have at our disposal a very well developed and mathematically strongly elaborated approach to contextual probability, because in the two decades that have passed we worked out – at that time purely for reasons of investigating the nature of the quantum probability model – an approach, which we called the 'hidden measurement approach', where the quantum probability model is defined starting from the idea of a contextual deviation of a classical probability model. Hence, the mathematical material that we developed for this hidden measurement approach will be fully used for a further elaboration of the contextual risk approach.